\RequirePackage{fix-cm} 
\documentclass[a4paper, twoside, reqno, dvips, 12pt]{amsart}
\usepackage{fixltx2e}   

\usepackage{etex}

\usepackage[latin1]{inputenc}
\usepackage[T1]{fontenc}

\usepackage{eucal}
\usepackage{esint}
\usepackage{dsfont}
\usepackage{xspace}
\usepackage{amsgen}
\usepackage{amsthm}
\usepackage{amssymb}
\usepackage{amsmath}
\usepackage{upgreek}
\usepackage{wasysym}
\usepackage{amsfonts}
\usepackage{stmaryrd}
\usepackage{mathtools}
\usepackage{textgreek}

\usepackage{mathrsfs}
\DeclareMathAlphabet{\mathscrbf}{OMS}{mdugm}{b}{n}

\usepackage{a4wide}

\usepackage[dvipsnames, table]{xcolor}
\definecolor{bckg}{RGB}{20.8, 20.8, 20.8}
\definecolor{oneblue}{rgb}{0.0, 0.0, 0.85}
\definecolor{Lightblue}{RGB}{214, 214, 214}
\definecolor{bluepigment}{rgb}{0.2, 0.2, 0.6}
\definecolor{charcoal}{rgb}{0.21, 0.27, 0.31}
\definecolor{denimblue}{rgb}{0.08, 0.38, 0.74}
\definecolor{Lightgray}{rgb}{0.89, 0.89, 0.89}
\definecolor{darkgrey}{rgb}{0.273, 0.281, 0.30}
\definecolor{darkelectricblue}{rgb}{0.33, 0.41, 0.47}

\usepackage{psfrag}
\usepackage{graphicx}
\usepackage{subfigure}
\usepackage{morefloats}
\usepackage{indentfirst}

\usepackage{acronym}
\usepackage{microtype}
\usepackage[labelsep=period,%
            labelfont={bf,sf,color=bluepigment},%
            justification=raggedright]{caption}

\usepackage[perpage, symbol]{footmisc}

\usepackage[usenames, dvipsnames]{pstricks}
\usepackage{epsfig}
\usepackage{pst-grad} 
\usepackage{pst-plot} 

\usepackage[colorlinks,
           urlcolor=oneblue,
           linkcolor=denimblue,
           citecolor=NavyBlue,
           bookmarksopen=false,
           pdfpagemode=UseNone,
           pagebackref]{hyperref}

\usepackage[sort&compress, comma, square, numbers]{natbib}

\usepackage[explicit]{titlesec}

\titleformat{\section}
  {\color{NavyBlue}\Large\sffamily\bfseries}
  {}
  {0em}
  {\colorbox{bckg!5}{\parbox{\dimexpr\linewidth-2\fboxsep\relax}{\centering\thesection. #1}}}
  [\vspace*{0.33em}]

\titleformat{name=\section,numberless}
  {\color{NavyBlue}\Large\sffamily\bfseries}
  {}
  {0.0em}
  {\colorbox{bckg!10}{\parbox{\dimexpr\linewidth-2\fboxsep\relax}{\centering#1}}}
  [\vspace*{0.33em}]

\titleformat{\subsection}
  {\color{NavyBlue}\large\sffamily\bfseries}
  {}
  {0.0em}
  {\colorbox{bckg!5}{\parbox{\dimexpr\linewidth-2\fboxsep\relax}{\centering\thesubsection. #1}}}
  [\vspace*{0.33em}]

\titleformat{name=\subsection,numberless}
  {\color{NavyBlue}\Large\sffamily\bfseries}
  {}
  {0em}
  {\colorbox{bckg!10}{\parbox{\dimexpr\linewidth-2\fboxsep\relax}{\centering#1}}}
  [\vspace*{0.33em}]

\titleformat{\subsubsection}
  {\color{bluepigment}\sffamily\normalsize\bfseries}
  {\thesubsubsection}
  {0.5em}
  {#1}
  [\vspace*{0.33em}]

\titleformat{\paragraph}[runin]
  {\color{bluepigment}\sffamily\small\bfseries}
  {}
  {0em}
  {#1}

\titlespacing{\section}{1.0em}{1.5em plus 2pt minus 2pt}%
{1.0em plus 2pt minus 2pt}[0em]
\titlespacing{\subsection}{1.0em}{1.5em plus 2pt minus 2pt}%
{1.0em}[0em]
\titlespacing{\subsubsection}{1.0em}{1.5em plus 2pt minus 2pt}%
{1.0em plus 2pt minus 2pt}[0em]

\usepackage{titletoc}

\contentsmargin{0.5em}
\newlength{\tocsep} 
\titlecontents{section}[\tocsep]
  {\addvspace{10pt}\bfseries\sffamily}
  {\contentslabel[\thecontentslabel]{\tocsep}}
  {}
  {\ \titlerule*[0.75pc]{.}\ \thecontentspage}
  []
\titlecontents{subsection}[\tocsep]
  {\addvspace{8pt}\sffamily}
  {\contentslabel[\thecontentslabel]{\tocsep}}
  {}
  {\ \titlerule*[0.5pc]{.}\ \thecontentspage}
  []
\titlecontents*{subsubsection}[\tocsep]
  {\addvspace{2pt}\footnotesize\sffamily}
  {}
  {}
  {\ \titlerule*[0.35pc]{.}\ \thecontentspage}
  [\\*]

\makeatletter
\def\@setauthors{%
  \begingroup
  \def\thanks{\protect\thanks@warning}%
  \trivlist
  \centering\footnotesize \@topsep30\p@\relax
  \advance\@topsep by -\baselineskip
  \item\relax
  \author@andify\authors
  \def\\{\protect\linebreak}%
  \textsc{\normalsize\textcolor{darkelectricblue}{\authors}}%
  \ifx\@empty\contribs
  \else
    ,\penalty-3 \space \@setcontribs
    \@closetoccontribs
  \fi
  \endtrivlist
  \endgroup
}
\def\@settitle{\begin{center}%
  \baselineskip14\p@\relax
    \bfseries
    \textsc{\Large\textcolor{charcoal}{\@title}}
  \end{center}%
}
\makeatother

\usepackage{enumitem}
\setlist[description]{%
  topsep=30pt,               
  itemsep=5pt,               
  font={\bfseries\sffamily\color{NavyBlue}}, 
}

\usepackage{fancyhdr}
\usepackage{lastpage}

\newcommand*\Title{\textcolor{bluepigment}{Regularisation shallow water equations}}
\newcommand*\Authors{\textcolor{bluepigment}{D.~Clamond \& D.~Dutykh}}
\newcommand*{\plogo}{\textcolor{gray}{{\texttt{arXiv.org} / \textsc{hal}}}} 

\numberwithin{equation}{section}

\newcommand{\U}{\mathcal{U}}
\newcommand{\ud}{\mathrm{d}}

\newcommand{\Rr}{\mathscr{R}}

\renewcommand{\alpha}{\upalpha}
\newcommand{\eps}{\upvarepsilon}

\renewcommand{\eta}{\text{\texteta}}
\renewcommand{\beta}{\text{\textbeta}}

\newcommand{\ie}{\emph{i.e.}\xspace}
\newcommand{\eg}{\emph{e.g.}\xspace}

\newcommand{\abs}[1]{\lvert\, #1\, \rvert}

\newcommand{\eqdef}{\mathop{\stackrel{\mathrm{def}}{=}}}

\newcommand{\half}{{\textstyle{1\over2}}}
\newcommand{\third}{{\textstyle{1\over3}}}

\newcommand{\sixth}{{\textstyle{1\over6}}}
\newcommand{\fourth}{{\textstyle{1\over4}}}
\newcommand{\twothird}{{\textstyle{2\over3}}}

\begin{document}

\title[\Title]{Non-dispersive conservative regularisation of nonlinear shallow water (and isothermal Euler) equations}

\author[D.~Clamond]{Didier Clamond$^*$}
\address{\textbf{D.~Clamond:} Universit\'e C\^ote d'Azur, Laboratoire J. A. Dieudonn\'e, CNRS UMR 7351, Parc Valrose, F-06108 Nice cedex 2, France}
\email{diderc@unice.fr}
\urladdr{http://math.unice.fr/~didierc/}
\thanks{$^*$ Corresponding author}

\author[D.~Dutykh]{Denys Dutykh}
\address{\textbf{D.~Dutykh:} LAMA, UMR 5127 CNRS, Universit\'e Savoie Mont Blanc, Campus Scientifique, 
F-73376 Le Bourget-du-Lac Cedex, France}
\email{Denys.Dutykh@univ-savoie.fr}
\urladdr{http://www.denys-dutykh.com/}



\begin{titlepage}
\thispagestyle{empty} 
\noindent
{\Large Didier \textsc{Clamond}}\\
{\it\textcolor{gray}{Universit\'e C\^ote d'Azur, France}}\\[0.02\textheight]
{\Large Denys \textsc{Dutykh}}\\
{\it\textcolor{gray}{CNRS, Universit\'e Savoie Mont Blanc, France}}
\\[0.16\textheight]

\colorbox{Lightblue}{
  \parbox[t]{1.0\textwidth}{
    \centering\huge\sc
    \vspace*{0.7cm}
    
    \textcolor{bluepigment}{Non-dispersive conservative regularisation of nonlinear shallow water (and isothermal Euler) equations}

    \vspace*{0.7cm}
  }
}

\vfill 

\raggedleft     
{\large \plogo} 
\end{titlepage}


\newpage
\thispagestyle{empty} 
\par\vspace*{\fill}   
\begin{flushright} 
{\textcolor{denimblue}{\textsc{Last modified:}} \today}
\end{flushright}


\newpage
\maketitle
\thispagestyle{empty}


\begin{abstract}

A new regularisation of the shallow water (and isentropic \textsc{Euler}) equations is proposed. The regularised equations are non-dissipative, non-dispersive and possess a variational structure. Thus, the mass, the momentum and the energy are conserved. Hence, for instance, regularised hydraulic jumps are smooth and non-oscillatory. Another particularly interesting feature of this regularisation is that smoothed `shocks' propagates at exactly the same speed as the original discontinuous ones. The performance of the new model is illustrated numerically on some dam-break test cases, which are classical in the hyperbolic realm.


\bigskip
\noindent \textbf{\keywordsname:} Shallow water flows; Dispersionless; Conservative; Regularisation. \\


\end{abstract}


\newpage
\tableofcontents
\thispagestyle{empty}


\clearpage
\section{Introduction}

\begin{figure}
\centering
\includegraphics[width=0.99\textwidth]{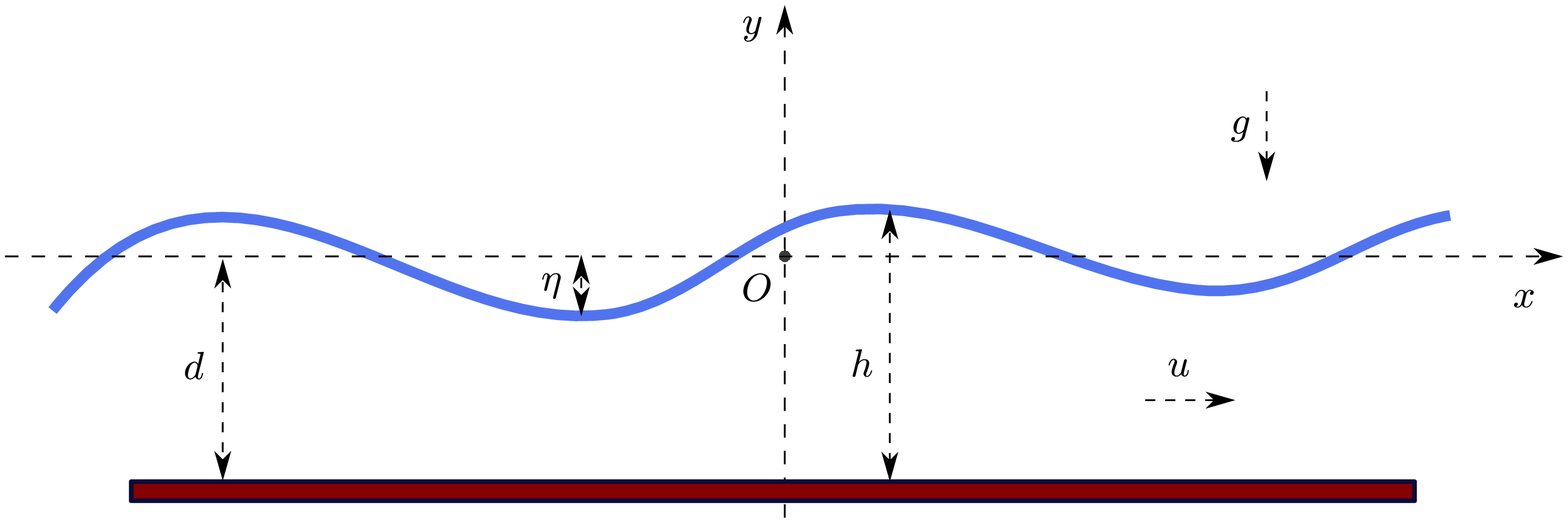}
\caption{\small\em Sketch of the fluid domain.}\label{fig:sketch}
\end{figure}

In fluid mechanics, many phenomena can be described by hyperbolic partial differential equations, such as the inviscid \textsc{Burgers} equation \cite{Burgers1948}, the isentropic \textsc{Euler} equations \cite{Majda2001} and the shallow water (\textsc{Airy} or \textsc{Saint-Venant} \cite{SV1871}) equations. The latter, for flat seabeds in one horizontal dimension, are most often written as mass and momentum flux conservations 
\begin{align}\label{eq:cSV1}
  h_t\ +\ \partial_x\!\left[\,h\,u\,\right]\ &=\ 0, \\
  \partial_t \left[\, h\,u\,\right]\, +\ \partial_x\!\left[\,h\,u^2\,+\,\half\,g\,h^2\,\right]\, &=\ 0,\label{eq:cSV2}
\end{align}
where $u\ =\ u(x,\,t)$ is the depth-averaged horizontal velocity ($x$ the horizontal coordinate, $t$ the time), $h\ =\ d\ +\ \eta\,(x,\,t)$ is the total water depth ($d$ the mean depth, $\eta$ the surface elevation from rest) and $g$ is the (downward) acceleration due to gravity (see the sketch in Figure~\ref{fig:sketch}). These equations describe nonlinear non-dispersive long surface gravity waves propagating in shallow water. They are equations of choice when one is interested in modelling large scale phenomena without resolving the details at the small scales, for instance, in tsunamis and tides modelling. It should be noted that equations (\ref{eq:cSV1}) -- (\ref{eq:cSV2}) are mathematically identical to the isothermal \textsc{Euler} equations describing some compressible fluids \cite{Chen2005a}. Here, we focus on shallow water equations, but it is clear that our claims apply as well to the isothermal \textsc{Euler} and mathematically similar equations.

Hyperbolicity is a nice feature of equations (\ref{eq:cSV1}) -- (\ref{eq:cSV2}) because they can be tackled analytically and numerically with powerful methods (\eg, characteristics, finite volumes, discontinuous \textsc{Galerkin}). A major inconvenient is that these equations admit non-unique weak solutions and entropy considerations have been proposed to ensure unicity \cite{Lax1973}. In gas dynamics, these weak solutions correspond to shock waves and the loss of regularity can be problematic, in particular for computations using spectral methods (even if some spectral approaches have been developed for hyperbolic equations as well \cite{Gottlieb2001a}). Several methods have then been introduced to regularise the equations and, in particular, to avoid the formation of sharp discontinuous shocks (replacing them by smooth $\tanh$-like profiles). Perhaps, the first regularisation was proposed by J.~\textsc{Leray} \cite{Leray1934} in the context of incompressible \textsc{Navier--Stokes} equations. His theoretical programme consisted in showing the existence of solutions in regularised equations, subsequently taking the limit $\eps\ \to\ 0$ ($\eps$ a small regularising parameter) to obtain weak solutions of \textsc{Navier--Stokes}.

A method of regularisation consists in first adding an artificial viscosity into 
the equations, and in taking the limit of vanishing viscosity in a second time. This method was introduced by \textsc{von Neumann} and \textsc{Richtmyer} \cite{VonNeumann1950}. It allows to generalise the classical concept of a solution and to prove eventually the uniqueness, existence and stability results for viscous regularised solutions \cite{Crandall1983, Bianchini2005}. Due to the added viscosity, the energy is no longer conserved, that can be a serious drawback for some applications, for instance for long time simulations when the shocks represent (unresolved) small scale phenomena that are not dissipative.

Another regularisation consists in adding weak dispersive effects to the equations \cite{Kondo2002a}. As shown by \textsc{Lax} \cite{Lax1983}, the dispersive regularisation is not always sufficient to obtain a reasonable limit to weak entropy solutions as the dispersion vanishes. Consequently, the most successful approach to study non-classical shock waves is to consider the combined dispersive-diffusive approximations \cite{Hayes2000}. Also, the added dispersion can generate high-frequency oscillations that must be resolved by the numerical scheme, resulting in a significant increase in the computational time. Nonlinear diffusive--dispersive regularisations for the scalar case were considered in \cite{Bedjaoui2004}. The main goal was to obtain a regularised model which admits the existence of classical solutions globally in time.

Yet another, less known, regularisation inspired by \textsc{Leray}'s method \cite{Leray1934}, consists filtered the velocity field such that the resulting equations are  non-dissipative and non-dispersive. Such regularisations have been proposed for the \textsc{Burgers} \cite{Bhat2006}, for isentropic \textsc{Euler} \cite{Bhat2007, Norgard2010a, Norgard2010} and other \cite{Camassa2010} equations. In the literature, this regularisation method appears with various denominations, such as {\em \textsc{Leray}-type regularisation}, {\em $\alpha-$regularisation} and {\em \textsc{Helmholtz} regularisation}. A drawback of this method is that the regularised (then smooth) shocks propagate at a speed different than that of the original equations. This drawback, among other things, is addressed in the present paper.

In this paper, we propose a new type of regularisation which is both non-dissipative and non-dispersive. This regularisation preserves the conservations of mass, momentum and energy, that is an important feature for some physical applications, in particular for long time simulations. The derivation proposed below is based on a recent work \cite{Clamond2015c} where a \textsc{Lagrangian}, suitable for long waves propagating in shallow water, was modified to incorporate one free parameter that can be used to improve the dispersion properties. Here, we make another step introducing two independent parameters. But, instead of improving the dispersion, these parameters are chosen to cancel the dispersion and thus to provide a regularisation of the classical shallow water (\textsc{Saint-Venant}) equations. In addition to being conservative and non-dispersive, this model yields regularised (smooth) `shocks' that propagate exactly at the same speed as in the original model. The properties of the obtained model are discussed below, mostly via numerical evidences.

The present manuscript is organised as follows. In Section~\ref{sec:model}, a new two-parameter generalised shallow water model is introduced. In Section~\ref{sec:RSWE}, the two parameters are related in a way to provide a non-dispersive non-diffusive and conservative regularised shallow water equations. The jump conditions of \textsc{Rankine--Hugoniot} type on both sides of a shock wave are discussed in Section~\ref{sec:rh}. These equations admit regular travelling wave solutions, as shown in Section~\ref{sec:steady}. In Section~\ref{sec:num}, several numerical examples are provided, demonstrating the efficiency of the method. Finally, some conclusions and perspectives are outlined in Section~\ref{sec:disc}.


\section{Model equations}
\label{sec:model}

For two-dimensional surface gravity waves propagating in shallow water over a horizontal seabed, the shallow water equations (\ref{eq:cSV1}) -- (\ref{eq:cSV2}) can be derived from the \textsc{Lagrangian} density
\begin{align}\label{defL0}
  \mathscr{L}_0\ \eqdef\ \half\,h\,u^2\ -\ \half\,g\,h^2\ +\ \left\{\,h_t\,+ \left[\,h\,u\,\right]_x\,\right\}\phi\,,
\end{align}
where $\phi\,(x,\,t)$ is, physically, a velocity ``potential'' introduced here as a \textsc{Lagrange} multiplier. Taking into account the next order of approximation, one can derive the very-well known fully-nonlinear weakly dispersive classical \textsc{Serre--Green--Naghdi} (cSGN) \cite{Green1974, Serre1953, Su1969, Wei1995, Wu2001a}. These equations can be derived in many ways, but the simplest derivation is from the \textsc{Euler--Lagrange} equations of the \textsc{Lagrangian} density
\begin{align}\label{defL1}
  \mathscr{L}_1\ \eqdef\ \half\,h\,u^2\ +\ \sixth\,h^3\,u_x^{\,2}\ -\ \half\,g\,h^2\ +\,\left\{\,h_t\,+\left[\,h\,u\,\right]_x\,\right\}\phi\,.
\end{align}
Recently, a modified version of these equations was proposed in \cite{Clamond2015c} in order to improve the dispersion properties, if needed. These improved \textsc{Serre--Green--Naghdi} (iSGN) can be derived from the \textsc{Lagrangian} density
\begin{align}\label{defL2}
  \mathscr{L}_2\ \eqdef\ \half\,h\,u^2\ +\,\left(\sixth\/+\/\fourth\/\beta\right)h^3\,u_x^{\,2}\ -\ \half\,g\,h^2\left(1\/+\/\half\/\beta\/h_{x}^{\,2}\right)\, +\,\left\{\,h_t\,+\left[\,h\,u\,\right]_x\,\right\}\phi,
\end{align}
where $\beta$ is a free parameter at our disposal. The \textsc{Lagrangian} densities $\mathscr{L}_1$ and $\mathscr{L}_2$ have the same order of approximation (see \cite{Clamond2015c} for details) and, obviously, $\mathscr{L}_1$ is a special case of $\mathscr{L}_2$ when $\beta\ =\ 0\,$. The case $\beta\ =\ 2/15$ is the best choice for improving the dispersion properties of infinitesimal waves. The reader is referred to \cite{Clamond2015c} for further details about the cSGN and iSGN models.

In the present paper, we consider a natural two-parameter extension of $\mathscr{L}_2$ in the form
\begin{align}\label{defL3}
  \mathscr{L}_3\ \eqdef\ \half\,h\,u^2\ +\,\left(\sixth\/+\/\fourth\/\beta_1\right)h^3\,u_x^{\,2}\ -\ \half\,g\,h^2\left(1\/+\/\half\/\beta_2\/h_{x}^{\,2}\right)\, +\ \left\{\,h_t\,+\left[\,h\,u\,\right]_x\,\right\}\phi\,.
\end{align}
This \textsc{Lagrangian} density obviously generalises the ones above. The corresponding \textsc{Euler--Lagrange} equations are
\begin{align}
  \delta\phi:\quad & 0\ =\ h_t\ +\,\left[\,h\,u\,\right]_x\,, \\
  \delta{u}:\quad & 0\ =\ h\,u\ +\ \phi\,h_x\ -\,\left[\,h\,\phi\,\right]_x\ -\ \left(\third\ +\ \half\,\beta_1\right)\!\left[\,h^3\,u_x\,\right]_x\,, \\
  \delta h:\quad & 0\ =\ \half\,u^2\ -\ g\,h\ -\ \phi_t\ +\ \phi\,u_x\ -\ [\,u\,\phi\,]_x \nonumber \\
  &\qquad+\,\left(\half\ +\ \textstyle{3\over4}\,\beta_1\right)h^2\,u_x^{\,2}\ -\ \half\,\beta_2\,g\,h\,h_{x}^{\,2}\,+\,\half\,\beta_2\,g\,[\,h^2\,h_x\,]_x\,,
\end{align}
thence one obtains the generalised \textsc{Serre--Green--Naghdi} (gSGN) equations
\begin{align}
  h_t\ +\ \partial_x\!\left[\,h\,u\,\right]\ &=\ 0\,, \label{GSWmass}\\
  \phi_{xt}\ +\ \partial_x\!\left[\,u\,\phi_x\,-\,\half\,u^2\,+\,g\,h\ -\ \left(\half+\textstyle{3\over4}\,\beta_1\right)h^2\,u_x^{\,2}\,-\,\half\,\beta_2\,g\,(\/h^2\/h_{xx}\/+\/h\/h_x^{\,2}\/)\,\right]\, &=\ 0\,,\label{GSWmomq}\\
  u\ -\ \left(\third\ +\ \half\,\beta_1\right)\!\left(\,3\,h\,h_x\,u_x\,+\,h^2\,u_{xx}\,\right)\ &=\ \phi_x\,. \label{GSWdefq}
\end{align}
From these equations, a non-conservative momentum equation can be obtained as
\begin{equation}\label{eq:ncons}
  u_t\ +\ u\,u_x\ +\ g\,h_x\ +\ \third\,h^{-1}\,\partial_x\!\left[\,h^2\,\Gamma\,\right]\ =\ 0\,,
\end{equation}
where 
\begin{equation*}
  \Gamma\ \eqdef\ \left(1\,+\,\textstyle{3\over2}\,\beta_1\right)h\left[\,u_x^{\,2}\, -\, u_{xt}\, -\, u\,u_{xx}\,\right]\, -\ \textstyle{3\over2}\,\beta_2\,g\left[\,h\,h_{xx}\,+\,\half\,h_x^{\,2}\,\right].
\end{equation*}
The quantity $\Gamma$ plays the role of a \emph{relaxed} version of fluid vertical acceleration at the free surface. Conservative equations for the momentum flux and the energy can also be obtained as
\begin{align}
  0\ &=\ \partial_t\!\left[\,h\,u\,\right]\, +\ \partial_x\!\left[\,h\,u^2\, +\, \half\,g\,h^2\, +\, \third\,h^2\,\Gamma\,\right], \label{GSWqdmflux1} \\
  0\ &=\ \partial_t\!\left[\,\half\,h\,u^2\,+\,(\sixth+\textstyle{1\over4}\/\beta_1)\,h^3\,u_x^{\,2}\,+\,\half\,g\,h^2\left(1+\textstyle{1\over2}\/\beta_2\/h_x^{\,2}\right)\right]\,+\ \partial_x\!\left[\left\{\half\,u^2\,+\right.\right.\nonumber\\
  &\qquad\left.\left.\ (\sixth+\textstyle{1\over4}\beta_1)\,h^2\,u_x^{\,2}\,+\,g\,h\left(1+\textstyle{1\over4}\beta_2h_x^{\,2}\right)+\,\third\,h\,\Gamma\,\right\}h\,u\,+\,\half\,\beta_2\,g\,h^3\,h_x\,u_x\,\right]. \label{enecons}
\end{align}
Thus, for any choice of the parameters $\beta_j\,$, the gSGN equations conserve mass, momentum and energy.

It should be noted that $\mathscr{L}_3$ is asymptotically consistent with $\mathscr{L}_1$ and $\mathscr{L}_2$ only if $\beta_1\ =\ \beta_2\,$. When $\beta_1\ \neq\ \beta_2\,$, $\mathscr{L}_3$ remains asymptotically consistent with $\mathscr{L}_0$ for all choices of the parameters $\beta_j\,$, however. Thus, in the next section, we exploit this feature to derive a regularised version of the \textsc{Saint-Venant} equations.


\section{Regularised shallow water equations}
\label{sec:RSWE}

Here, we consider non-dispersive version of the gSGN equations above, obtaining thus a conservative regularised modification of the classical (\ie dispersionless) shallow water (\textsc{Saint-Venant}) equations.

\subsection{Linear dispersion relation}

For infinitesimal waves, $\eta$ and $u$ being small, the gSGN equations can be linearised as 
\begin{align*}
  \eta_t\ +\ d\,u_x\ &=\ 0\,, \qquad
  u_t\ -\,\left(\third\,+\,\half\,\beta_1\right)d^2\,u_{\,xxt}\ +\ g\,\eta_{\,x}\ -\ \half\,\beta_2\,g\,d^{\,2}\,\eta_{\,xxx}\ =\ 0\,.
\end{align*}
Seeking for travelling wave solutions of the form $\eta\ =\ a\,\cos k\,(x\ -\ c_0\,t)\,$, one obtains the linear dispersion relation 
\begin{equation} \label{eq:dispRel0}
  \frac{c_0^{\,2}}{g\,d}\ =\ \frac{2\ +\ \beta_2\,(k\,d)^2}{2\,+\,(\twothird\,+\,\beta_1)\,(k\,d)^2}\,.
\end{equation}
A non dispersive model (\ie $c_0$ independent of $k$) is obtained taking $\beta_1\ =\ \beta_2\ -\ 2/3\,$. It is this special choice that is investigated in this paper and that we call the {\em regularised} \textsc{Saint-Venant} (rSV) equations.


\subsection{Regularised Saint-Venant (rSV) equations}

Choosing $\beta_1\ \eqdef\ 2\/\eps\ -\ 2/3$ and $\beta_2\ \eqdef\ 2\/\eps$ (in order to obtain a non-dispersive model) and introducing $\Gamma\ \eqdef\ 3\/\eps\/\Rr$ for convenience ($\eps$ being a free parameter), the \textsc{Lagrangian} density $\mathscr{L}_3$ becomes 
\begin{align}\label{defLeps}
  \mathscr{L}_\epsilon\ \eqdef\ \half\,h\,u^2\ -\ \half\,g\,h^2\  +\,\left\{\,h_t\, + \left[\,h\,u\,\right]_x\,\right\}\phi\ +\ \half\,\epsilon\,h^2\left(\,h\,u_x^{\,2}\, -\, g\,h_{x}^{\,2}\right)\,,
\end{align}
and the resulting equations are
\begin{align}
  h_t\ +\ \partial_x\!\left[\,h\,u\,\right]\ &=\ 0\,, \label{NDGSEmass}\\
  \partial_t\!\left[\,h\,u\,\right]\, +\ \partial_x\!\left[\,h\,u^2\,+\,\half\,g\,h^2\,+\,\eps\,\Rr\,h^2\,\right]\ &=\ 0\,, \label{NDGSEmomG}\\
  h\left(\,u_x^{\,2}\,-\,u_{xt}\,-\,u\,u_{xx}\,\right)\, -\ g\left(\,h\,h_{xx}\,+\,\half\,h_x^{\,2}\,\right)\, &\eqdef\ \Rr\,. \label{NDGSEdefG}
\end{align}
If $\eps\ =\ 0\,$, the classical \textsc{Saint-Venant} (cSV), or nonlinear shallow water equations (NSWE), are recovered. For $\eps\ \neq\ 0\,$, $\Rr$ is a regularising term that prevents the formation of shocks, as shown below in Section~\ref{sec:num}. Of course, the rSV equations yield a conservative equation for the energy
\begin{align}
  \partial_t\!\left[\,\half\,h\,u^2\,+\,\half\,g\,h^2\,+\,\half\,\epsilon\,h^3\,u_x^{\,2}\,+\,\half\,\epsilon\,g\,h^2\,h_x^{\,2}\,\right]\,+&\nonumber\\
  \partial_x\!\left[\left\{\,\half\,u^2\,+\,g\,h\,+\,\half\,\epsilon\,h^2\,u_x^{\,2}\,+\,\half\,\epsilon\,g\,h\,h_x^{\,2}\,+\,\epsilon\,h\,\Rr\,\right\}h\,u\,+\,\eps\,g\,h^3\,h_x\,u_x\,\right]& =\ 0\,. \label{NDGSEene}
\end{align} 
It should be noted that, though $\Rr$ involves high-order derivatives, the resulting rSV equations are not dissipative and not dispersive. For numerical resolutions, (\ref{NDGSEmomG}) can be advantageously replaced by one of the two equations
\begin{align}
  \partial_t\!\left[\,h\,\U\,\right]\, +\ \partial_x\!\left[\,h\,u\,\U\,+\,\half\,g\,h^2\, -\, \eps\,h^2\left(\,2\,h\,u_x^{\,2}\,+\,g\,h\,h_{xx}\,+\,\half\,g\,h_x^{\,2}\,\right)\,\right]\, &=\ 0\,, \label{GSWqdmflux2} \\
  \U_t\ +\ \partial_x\!\left[\,u\,\U\,-\,\half\,u^2\,+\,g\,h\,-\,\textstyle{3\over2}\,\eps\,h^2\,u_x^{\,2}\,-\,\eps\,g\,h\left(\,h\,h_{xx}\,+\,h_x^{\,2}\,\right)\,\right]\, &=\ 0\,, \label{RSWEmomq}
\end{align}
where the new variable $\U\/\eqdef\/u\/-\/\eps\/h\/(\/3\/h_x\/u_x\/+\/h\/
u_{xx}\/)$ can be physically interpreted as an approximation of the tangential 
velocity at the free surface (see \cite[Appendix~B]{Clamond2015c}). Note 
that $h\U\ =\ hu\ -\ \eps\,\partial_x[h^3\,u_x]$ and that rSV is different from the regularised isentropic \textsc{Euler} equations proposed in \cite{Bhat2007} where both the mass and momentum equations are modified.


\subsection{Steady motion}
\label{sec:steady}

Consider now the special case of travelling waves with permanent form, studied in the frame of reference moving with the wave where the flow is steady. The functions $h$ and $u$ being then independent of the time $t\,$, the mass conservation can be integrated as
\begin{equation}\label{solmasste}
  u\,h\ =\ \text{constant}\ \eqdef\ -\,c\,d \qquad \Longrightarrow\qquad u\ =\ -\,c\,d\,/\,h\,,
\end{equation}
so $c$ is the wave phase velocity observed in the frame of reference without mean flow. In the latter frame of reference, the wave travels toward the increasing $x-$direction if $c\ >\ 0\,$. With (\ref{solmasste}) and after some algebra, equations (\ref{GSWqdmflux2}) and (\ref{RSWEmomq}) give, respectively,
\begin{align}
  \frac{2\,\epsilon\,(\,h\,h_{\,xx}\,-\,h_{\,x}^{\,2}\,)}{g\,h\,/\,c^{\,2}}\ -\ \frac{\epsilon\,(\,2\,h\,h_{\,xx}\,+\,h_{\,x}^{\,2}\,)}{d^2\,/\,h^2}\ +\ \frac{2\,c^{\,2}}{g\,h}\ +\ \frac{h^2}{d^{\,2}}\ &=\ 1\ +\ \frac{2\,c^{\,2}}{g\,d}\ +\ \mathcal{C}_1\,,  \label{eqqdmbisseperm}\\
  \frac{\epsilon\,(\,2\,h\,h_{\,xx}\,-\,h_{\,x}^{\,2}\,)}{2\,g\,h^2\,/\,c^{\,2}\,d}\ -\ \frac{\epsilon\,(\,h\,h_{\,xx}\,+\,h_{\,x}^{\,2}\,)}{d\,/\,h}\ +\ \frac{c^{\,2}\,d}{2\,g\,h^2}\ +\ \frac{h}{d}\ &=\ 1\ +\ \frac{c^{\,2}}{2\,g\,d}\ +\ \mathcal{C}_2\,, \label{eqqdmfluxseperm}
\end{align}
where $\mathcal{C}_j$ are dimensionless integration constants ($\mathcal{C}_1\ =\ \mathcal{C}_2\ =\ 0$ if $h\ \to\ d$ as $x\ \to\ \infty$). Eliminating $h_{xx}$ between these two relations, one obtains the first-order ordinary differential equation
\begin{align}\label{RSWEODE}
  \eps\left(\frac{\ud\,h}{\ud\/x}\right)^{\!2}\ =\ \frac{\mathcal{F}\,-\,(1+\mathcal{C}_1 + 2\mathcal{F})\,(h/d)\,+\,(2+2\mathcal{C}_2+\mathcal{F})\,(h/d)^2\,-\,(h/d)^3}{\mathcal{F}\,-\,(h/d)^3}\,,
\end{align}
where $\mathcal{F}\ \eqdef\ c^2/gd$ is a \textsc{Froude} number squared. If $\eps\ =\ 0\,$, the equation (\ref{RSWEODE}) does not admit physically admissible regular smooth solutions.

If $\eps\ \neq\ 0\,$, exact solutions can be easily obtained in parametric form $x\ =\ x(\xi)\,$, $h\ =\ h(\xi)\ =\ d\ +\ \eta\,(\xi)$ and in terms of \textsc{Jacobian} elliptic functions. For brevity, we give here only the solitary wave (\ie $\mathcal{C}_1\ =\ \mathcal{C}_2\ =\ 0$) solution
\begin{equation}\label{solitonRSVE}
  x\, =\/ \int_0^\xi\left[\,\frac{\mathcal{F}\/d^3\,-\,h^3(\xi')}{(\mathcal{F}-1)\,d^3}\,\right]^{1\over2}\/\ud\/\xi', \quad \frac{\eta(\xi)}{d}\, =\, (\mathcal{F}-1)\,\operatorname{sech}^2\!\left(\kappa\,\xi\right),  \quad (\kappa\/d)^2\, =\, \frac{1}{\eps}.
\end{equation}
This solution is admissible only if $\eps$ is positive. It corresponds to a dispersionless solitary wave, as can be seen in the independence of the trend parameter $\kappa$ with respect of the amplitude. This is therefore not a suitable model for solitary surface gravity waves. However, is some media, there exist dispersionless solitary waves \cite{Nesterenko2001} and the rSV equations could be used as model (with, of course, different physical interpretations of the variables $h$ ans $u$ and of the parameters $g$ and $d$).

Note that the solitary wave solution shows that the phase velocity varies with the wave amplitude. The rSV waves (like the cSV ones) have thus {\em amplitude dispersion} though they have no {\em frequency dispersion}. In this paper {\em dispersion} always refers to {\em frequency dispersion}.


\subsection{Rankine--Hugoniot type conditions}
\label{sec:rh}

In the theory and practice of hyperbolic equations, it is well known that equations in the velocity $u$ or $\U$ such as (\ref{RSWEmomq}) with $\eps\ \to\ 0$ are not suitable for discontinuous solutions since they yield physically incorrect \textsc{Rankine--Hugoniot} jump conditions \cite{Godlewski1990, Godlewski1996}. However, this is not a problem if $\eps\ \neq\ 0$ because no (discontinuous) shocks are formed, as illustrated numerically below (see Section~\ref{sec:num}).

Assuming that $h_x$ and $u_x$ are both continuous if $\eps\ >\ 0$ and that discontinuities (if any) occur only in $h_{xx}$ and $u_{xx}$ (and thus in $\U$ too). Equations (\ref{GSWqdmflux2}) and (\ref{RSWEmomq}) (together with $\llbracket\/\U\/\rrbracket\ =\ -\eps\/h^2\llbracket\/u_{xx}\/\rrbracket$) yield the same jump condition
\begin{equation}\label{jumpqdm}
  \left(u\,-\,\dot{s}\right)\llbracket\,u_{xx}\,\rrbracket\, +\ g\,\llbracket\,h_{xx}\,\rrbracket\, =\ 0\,,
\end{equation}
where $\dot{s}\ \eqdef\ \ud s/\ud t$ is the speed of the smoothed shock located at $x\ =\ s\,(t)$ and $\llbracket\/f\/\rrbracket\eqdef f(x\!=\!s^+)-f(x\!=\!s^-)$ denotes the jump across the shock for any function $f\,$. For brevity, we call `shock' both the discontinuous (classical) and smoothed (regularised) shocks.

Differentiating twice with respect of $x$ the mass conservation (\ref{NDGSEmass}), the jump condition of the resulting equation is 
\begin{equation}\label{jumpmass}
  \left(u\,-\,\dot{s}\right)\llbracket\,h_{\,xx}\,\rrbracket\,+\,h\,\llbracket\,u_{\,xx}\,\rrbracket\, =\ 0\,,
\end{equation}
and the elimination of $\llbracket u_{xx}\rrbracket$ (or $\llbracket h_{xx}\rrbracket$) between (\ref{jumpqdm}) and (\ref{jumpmass}) yields at once 
\begin{equation*}
  \dot{s}(t)\ =\ u(x,\,t)\ \pm\ \sqrt{\,g\,h(x,t)\,} \qquad \text{ at }\quad x\,=\,s(t)\,.
\end{equation*}
The shock speed is thus independent of $\eps$ and the shock propagates along the characteristic lines of the classical \textsc{Saint-Venant} equations.

Under the same regularity conditions, the jump condition for the momentum and energy equations, respectively (\ref{NDGSEmomG}) and (\ref{NDGSEene}), both yield  $\llbracket\/\Rr\/\rrbracket\ =\ 0$ (\ie, $\Rr$ is continuous), thus from (\ref{NDGSEdefG}) 
\begin{equation} \label{jumpene}
  \llbracket\,u_{xt}\,\rrbracket\, +\ u\,\llbracket\,u_{xx}\,\rrbracket\, +\ g\,\llbracket\,h_{xx}\,\rrbracket\, =\ 0\,,
\end{equation}
thence, using \eqref{jumpqdm}, one gets
\begin{equation}\label{jumpuxt}
  \llbracket\,u_{xt}\,\rrbracket\ =\ -\,\dot{s}\,\llbracket\,u_{xx}\,\rrbracket\,.
\end{equation}
The continuity of $\Rr$ is compatible with its definition (\ref{NDGSEdefG}) and with (\ref{jumpqdm}) and (\ref{jumpuxt}), \ie
\begin{equation*}
  -\/\llbracket\/\Rr\/\rrbracket\, =\ h\,\llbracket\/u_{xt}\/\rrbracket\, +\ h\,u\,\llbracket\/u_{xx}\/\rrbracket +\, g\,h\,\llbracket\/h_{xx}\/\rrbracket\ =\ h\left\{\,(u\/-\/\dot{s})\,\llbracket\/u_{xx}\/\rrbracket\,+\, g\,\llbracket\/h_{xx}\/\rrbracket\,\right\}\,=\ 0\,.
\end{equation*}

$\Rr$ being continuous across the shock, applying one spatial derivative $\partial_x$ to the momentum equation (\ref{NDGSEmomG}), the jump condition for the resulting equation is $\llbracket\/\Rr_x\/\rrbracket = 0\,$, so $\Rr_x$ too is continuous across the shock. Applying the same procedure to the energy equation (\ref{NDGSEene}), one gets the jump condition
\begin{equation*}
  \left(u-\dot{s}\right)\left(\,h\,u_x\,\llbracket\/u_{\,xx}\/\rrbracket\,+\,g\,h_x\,\llbracket\/h_{xx}\/\rrbracket\,\right)\, +\ g\,h\left(\,h_x\,\llbracket\/u_{\,xx}\/\rrbracket\,+\,u_x\,\llbracket\/h_{xx}\/\rrbracket\,\right)\ =\ 0\,,
\end{equation*}
which, with (\ref{jumpqdm}) and (\ref{jumpmass}), is identically fulfilled. Thus, the derivative of the energy equation does not provide any additional information. It should be noted that all the jump relations above are independent of the regularising parameter $\eps\,$. Moreover, the regularising term vanishes identically on constant states; this property is necessary to preserve shock conditions of the original hyperbolic system.


\subsection{Remarks on the total energy}

For a domain $\Omega\,$, the energy density $\mathscr{H}_\epsilon$ and total energy $\mathcal{H}_\epsilon$ of the rSV equation are
\begin{equation}
  \mathscr{H}_\epsilon(x,t)\ \eqdef\ \half\,h\,u^2\,+\,\half\,g\,h^2\,+\,\half\,\epsilon\,h^3\,u_x^{\,2}\,+\,\half\,\epsilon\,g\,h^2\,h_x^{\,2}\,, \qquad \mathcal{H}_\epsilon(t)\ \eqdef\ \int_\Omega\mathscr{H}_\epsilon(x,t)\,\ud\/x\,.
\end{equation}
If $\Omega$ is periodic or if the flux of energy is constant at the boundaries of $\Omega\,$, then $\mathcal{H}_\epsilon$ is constant (\ie, $\ud\mathcal{H}_\epsilon/\ud\/t\ =\ 0$) because the energy equation (\ref{NDGSEene}) is conservative. However, the quantity $\mathcal{H}_0\,(t)$ (corresponding to the energy of the cSV equations) is not constant, in general. 

If $\mathscr{H}_0$ represents the density of physical energy (kinetic plus potential) then $\mathscr{H}_\epsilon\ -\ \mathscr{H}_0$ can be interpreted as a density of `internal' energy, $\mathcal{H}_\epsilon\ -\ \mathcal{H}_0$ being the `internal' energy of the domain $\Omega\,$. This `internal' energy can also be interpreted as an `entropy'. Indeed, when the temporal evolution of a smooth initial condition leads to stiffening of the free surface (see numerical simulations below), the quantities $\abs{h_x}$ and $\abs{u_x}$ increase in time in the vicinity of the forming shock, and so does $\mathscr{H}_\epsilon\ -\ \mathscr{H}_0\,$. We have then a transfert of energy between the physical one to the `internal' one. This behaviour is consistent with the cSV equations where the energy decreases across shocks.


\section{Numerical illustrations}
\label{sec:num}

In order to study the properties of the proposed regularisation method, we solve numerically the equations (\ref{NDGSEmass}) -- (\ref{NDGSEdefG}) using a \textsc{Fourier}-type pseudo-spectral method. It is totally fine since for any $\eps\ >\ 0$ we are dealing with smooth solutions. The periodicity is enforced by symmetrising the solution. The time stepping is done with automatic time step selection using embedded \textsc{Runge--Kutta} methods \cite{Shampine1997}. The relative and absolute errors were set to $10^{-10}\,$. Our pseudo-spectral approach to non-hydrostatic equations is described in \cite{Dutykh2011a}. As anti-aliasing we use an eight-order Erfc--Log filter \cite{Boyd1995}. In order to compare the regularised solutions with entropic solutions to the original hyperbolic system, we employ the finite volume scheme described in some detail in \cite{Dutykh2013}. Thus, we do not reproduce the numerical details in the present study. All numerical results are performed in dimensionless variables where $g\ =\ d\ =\ 1\,$.


\begin{figure}
  \centering
  \subfigure[$t\,\sqrt{g/d}\, =\, 0$]{\includegraphics[width=0.91\textwidth]{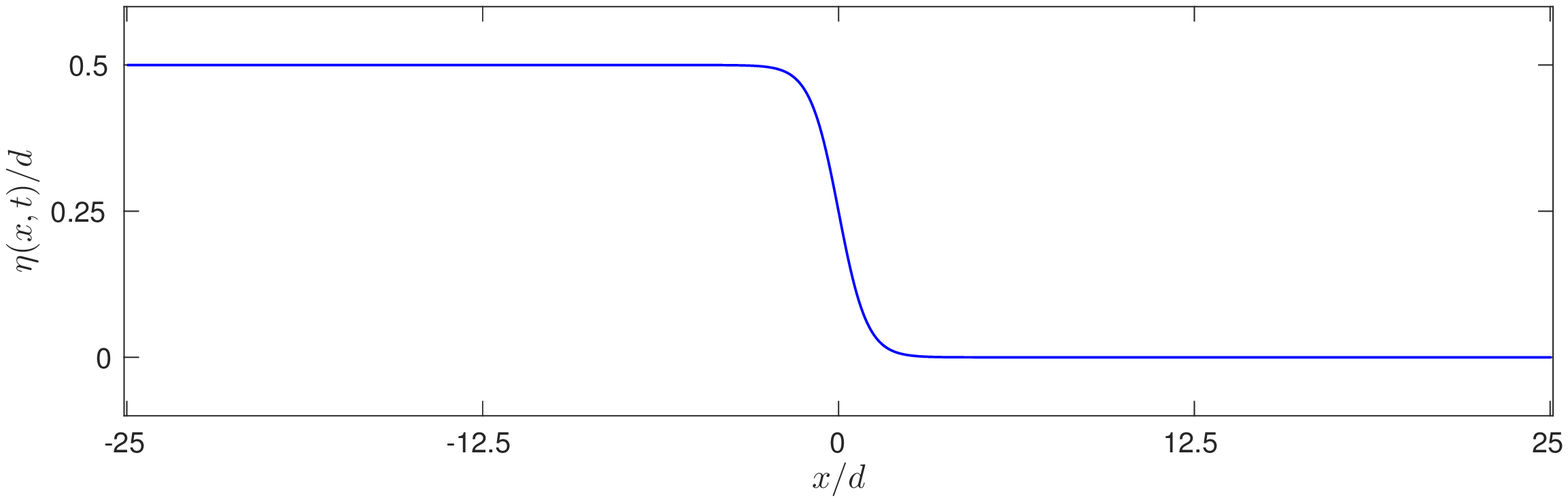}}
  \subfigure[$t\,\sqrt{g/d}\, =\, 5$]{\includegraphics[width=0.91\textwidth]{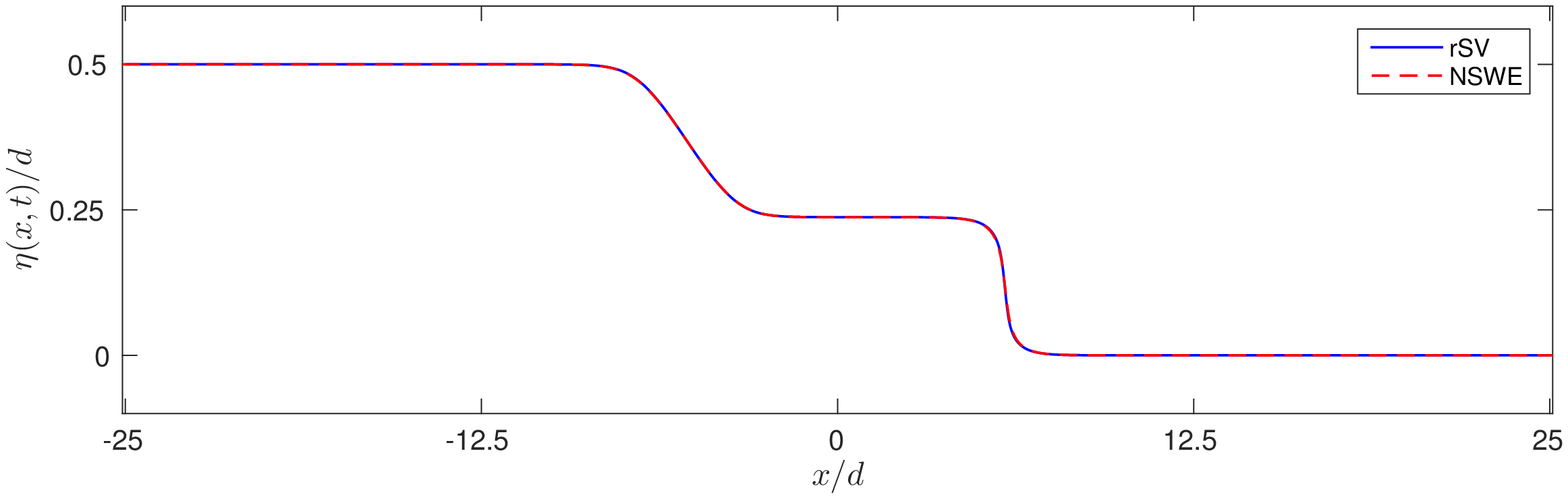}}
  \subfigure[$t\,\sqrt{g/d}\, =\, 10$]{\includegraphics[width=0.91\textwidth]{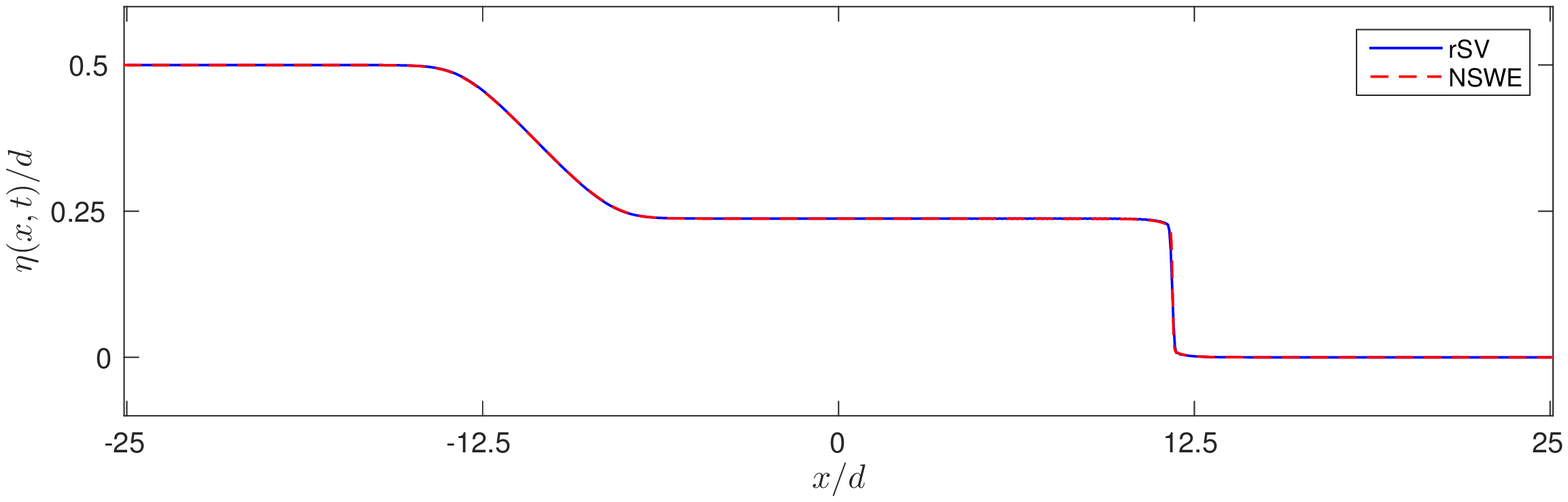}}
  \subfigure[$t\,\sqrt{g/d}\, =\, 15$]{\includegraphics[width=0.91\textwidth]{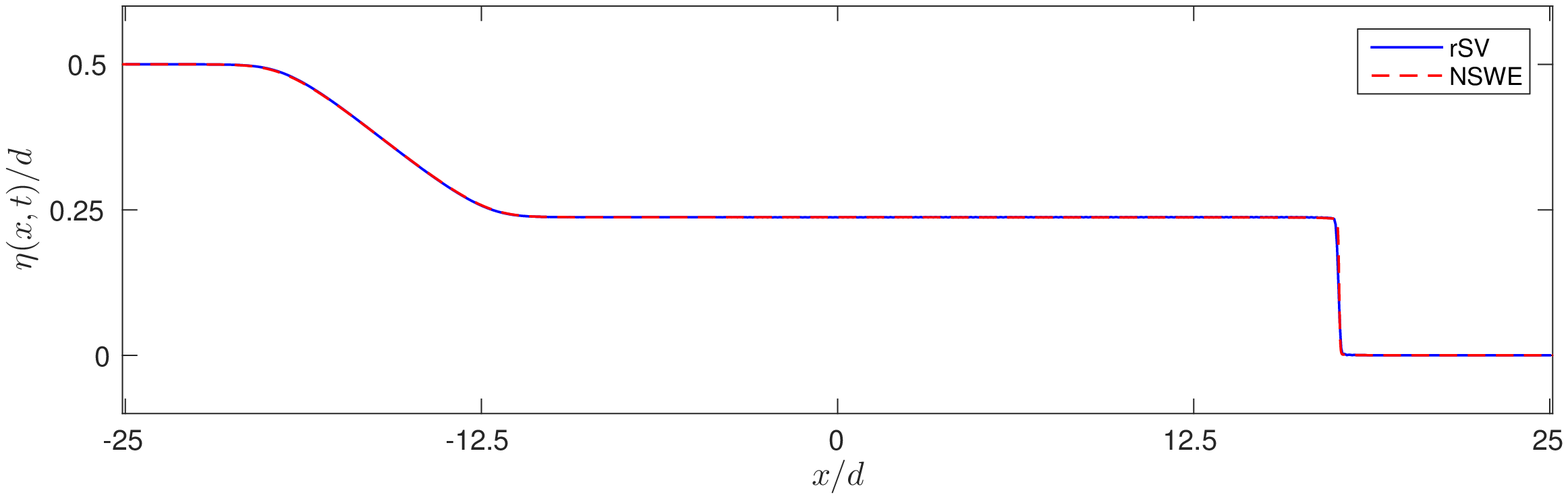}}
  \caption{\small\em Dam break problem solved with the hyperbolic and regularised models. The common initial condition is shown on the upper panel (regularisation parameter $\eps\ =\ 10^{-3}$).}
  \label{fig:dam}
\end{figure}

\begin{figure}
  \centering
  \subfigure[$t\,\sqrt{g/d}\, =\, 5$]{\includegraphics[width=0.91\textwidth]{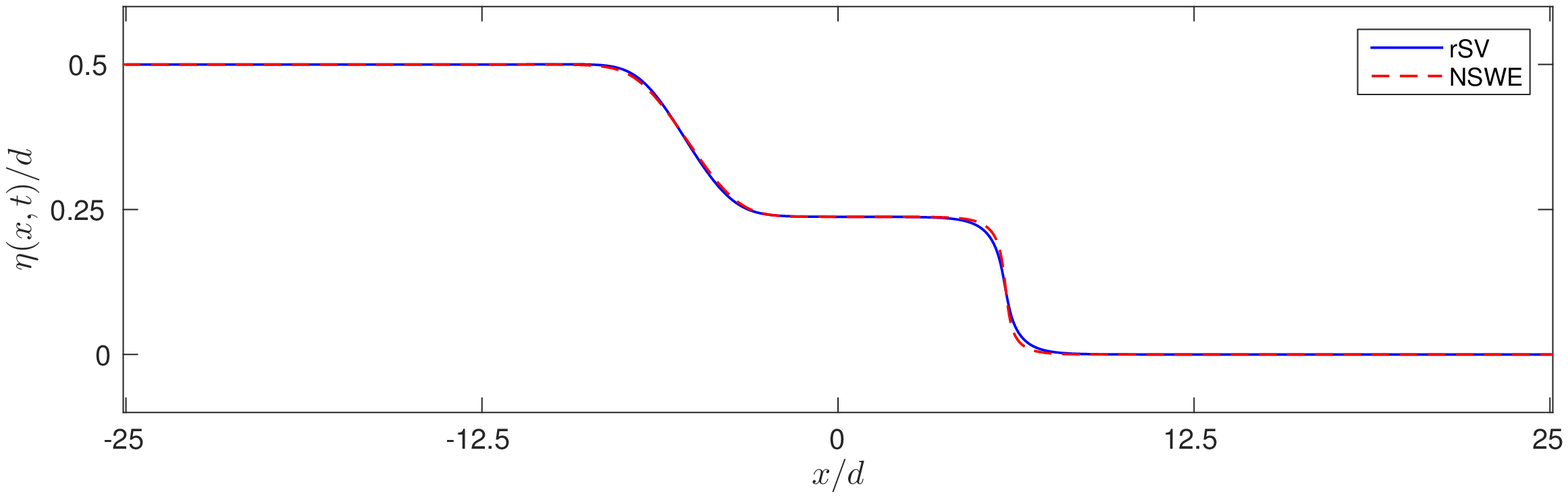}}
  \subfigure[$t\,\sqrt{g/d}\, =\, 10$]{\includegraphics[width=0.91\textwidth]{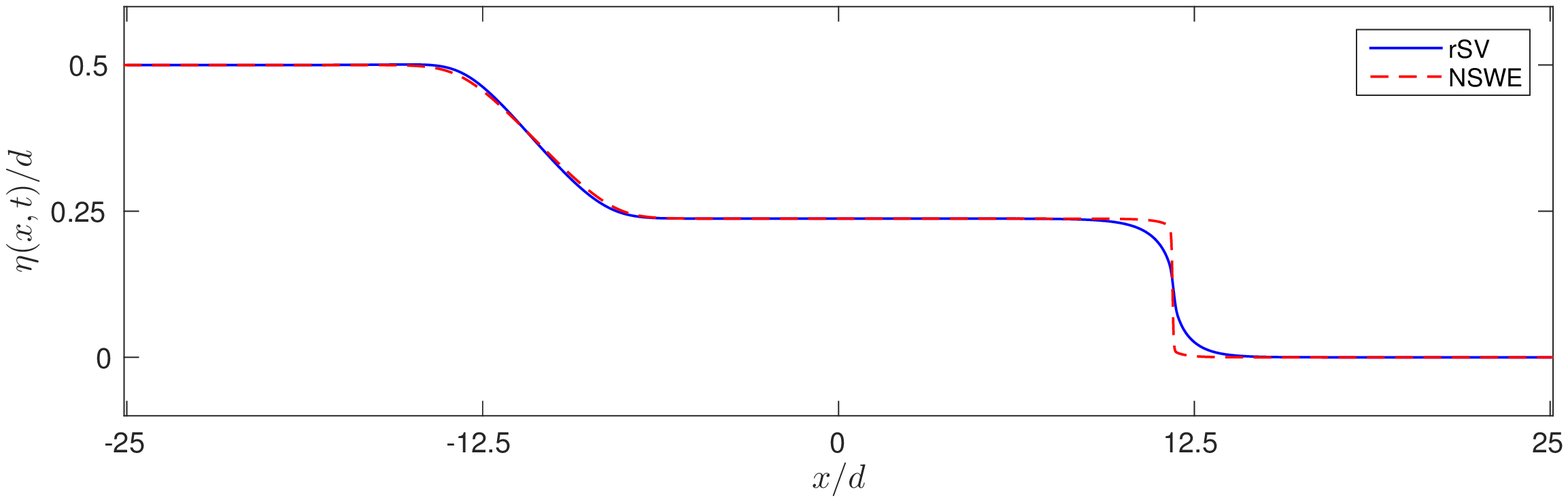}}
  \subfigure[$t\,\sqrt{g/d}\, =\, 15$]{\includegraphics[width=0.91\textwidth]{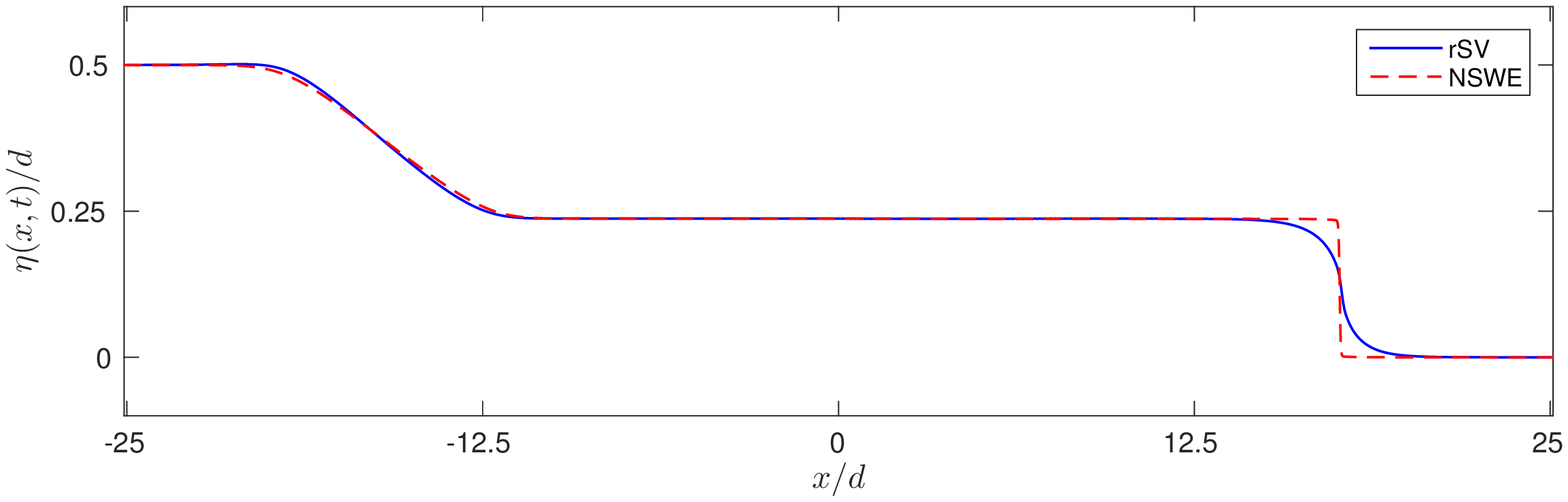}}
  \caption{\small\em Same problem as Figure~\ref{fig:dam} with $\eps\ =\ 1\,$.}
  \label{fig:dam2}
\end{figure}

\begin{figure}
  \centering
  \subfigure[$t\,\sqrt{g/d}\, =\, 5$]{\includegraphics[width=0.91\textwidth]{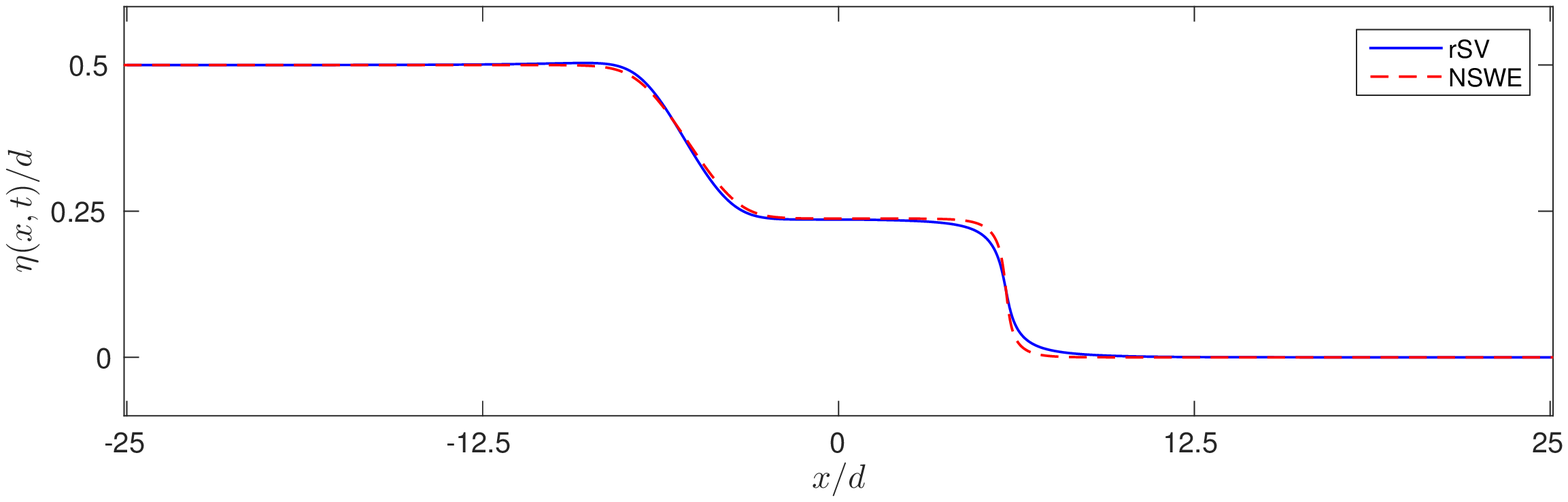}}
  \subfigure[$t\,\sqrt{g/d}\, =\, 10$]{\includegraphics[width=0.91\textwidth]{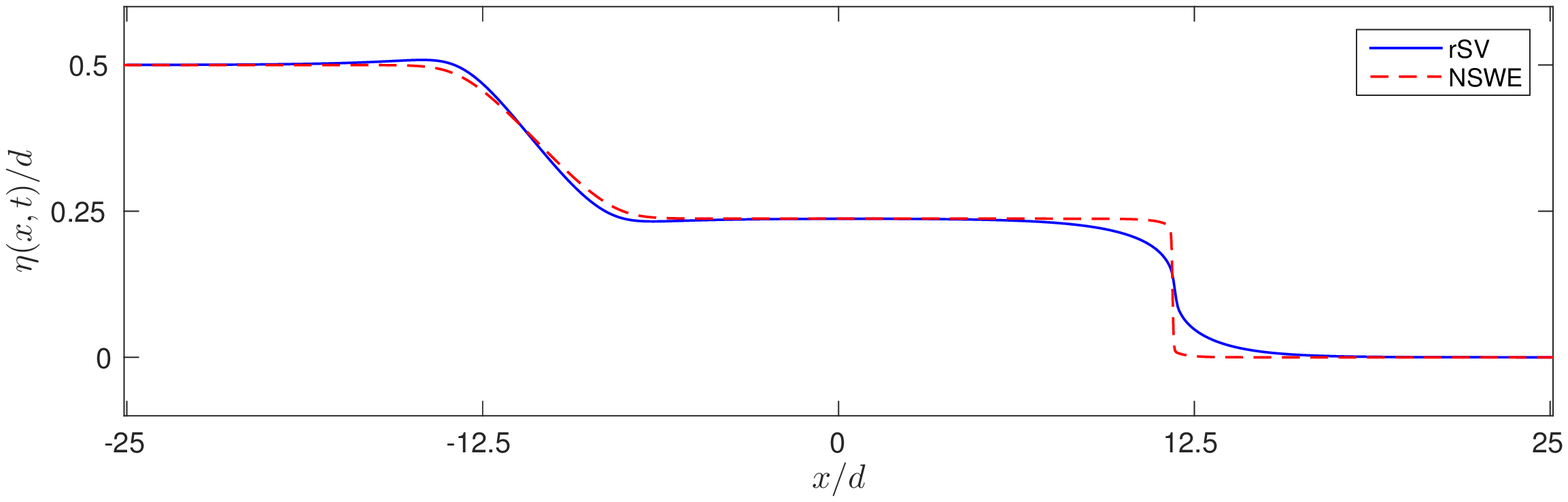}}
  \subfigure[$t\,\sqrt{g/d}\, =\, 15$]{\includegraphics[width=0.91\textwidth]{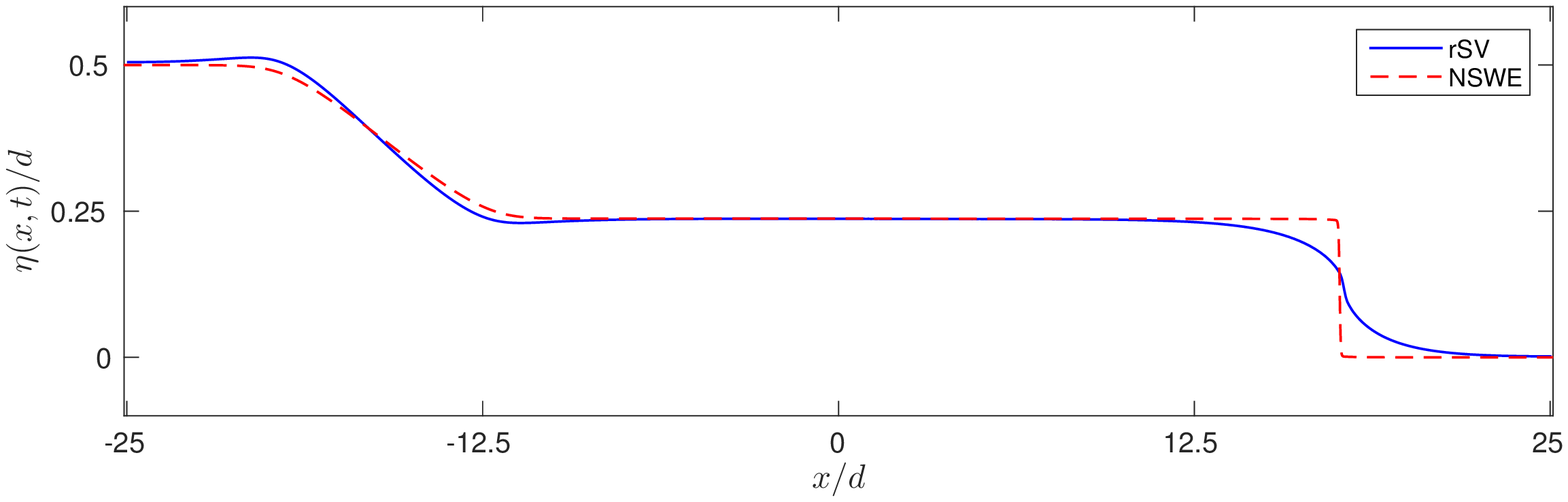}}
  \caption{\small\em Same problem as Figure~\ref{fig:dam} with $\eps\ =\ 5\,$.}
  \label{fig:dam3}
\end{figure}


\subsection{Dam-break problem}

The dam-break problem has become the standard test-case for shallow water equations \cite{Dutykh2010c}. Here, we consider the so-called wet dam-break problem where the water depth is positive everywhere and a wave system is generated due to the initial difference of water level on the left and on the right from a certain point, which is chosen to be the origin without loss of generality. The initial condition is chosen as a regularised wave front:
\begin{align*}
  h_0(x)\ &=\ h_\mathrm{l}\ +\ \half\left(h_\mathrm{r}-h_\mathrm{l}\right)\left(1+ \tanh(\delta\/x)\right), \\
  u_0(x)\ &=\ u_\mathrm{l}\ +\ \half\left(u_\mathrm{r}-u_\mathrm{l}\right)\left(1+ \tanh(\delta\/x)\right).
\end{align*}
The initial condition parameters are given in Table~\ref{tab:dam} (central column) and it is depicted in Figure~\ref{fig:dam}\textit{a}. The temporal evolution of this initial condition is shown in Figure~\ref{fig:dam}\textit{b}--\textit{d}. In particular, one can see that during the propagation, the initially smooth front becomes steeper to produce a real shock wave propagating rightwards. On the left, we observe a rarefaction wave according to the classical \textsc{Riemann} problem solution. The most important point to notice is that we have no dissipation and no dispersion in the regularised solution according to our model construction. For the value of the regularisation parameter $\eps\ =\ 10^{-3}$ the curves are non-distinguishable to graphical resolution.

\begin{table}
  \centering
  \begin{tabular}{l|c|c}
  \hline\hline
  \textit{Parameter} & \textit{Dam-break value} & \textit{Shock wave value} \\
  \hline\hline
  Regularisation parameter, $\epsilon$ & $10^{-3}$ & $10^{-2}$ \\
  Domain half-length, $\ell/d$ & $25$ & $25$ \\
  Number of \textsc{Fourier} modes, $N$ & $1024$ & $1024$ \\
  Initial transition length, $1/\delta d$ & $1$ & $1$ \\
  Final simulation time, $T\sqrt{g/d}$ & $15$ & $5$ \\
  Water depth on the left, $h_\mathrm{l}/d$ & $1.5$ & $1.5$ \\
  Water depth on the right, $h_\mathrm{r}/d$ & $1$ & $1$ \\
  Horizontal velocity on the left, $u_\mathrm{l}/\sqrt{gd}$ & $0$ & $1$ \\
  Horizontal velocity on the right, $u_\mathrm{r}/\sqrt{gd}$ & $0$ & $0.5435645\ldots$ \\
  \hline\hline
  \end{tabular}
  \bigskip
  \caption{\small\em Numerical and physical parameters used in the dam-break and shock wave simulations.}
  \label{tab:dam}
\end{table}

An important feature of the rSV equations is that the smoothed shock speed is independent of the regularisation parameter $\eps\,$. This can be clearly seen doing the same simulations as in Figure~\ref{fig:dam}, but with $\eps\ =\ 1$ (Fig.~\ref{fig:dam2}) and with $\eps\ =\ 5$ (Fig.~\ref{fig:dam3}). In these two simulations, the discrepancies between the regularised and original shallow water equations are of course more important than with small $\eps\ =\ 10^{-3}\,$, but it is clear that the shock speeds are not affected by the regularisation.


\subsection{Shock wave}

\begin{figure}
\centering
\includegraphics[width=0.99\textwidth]{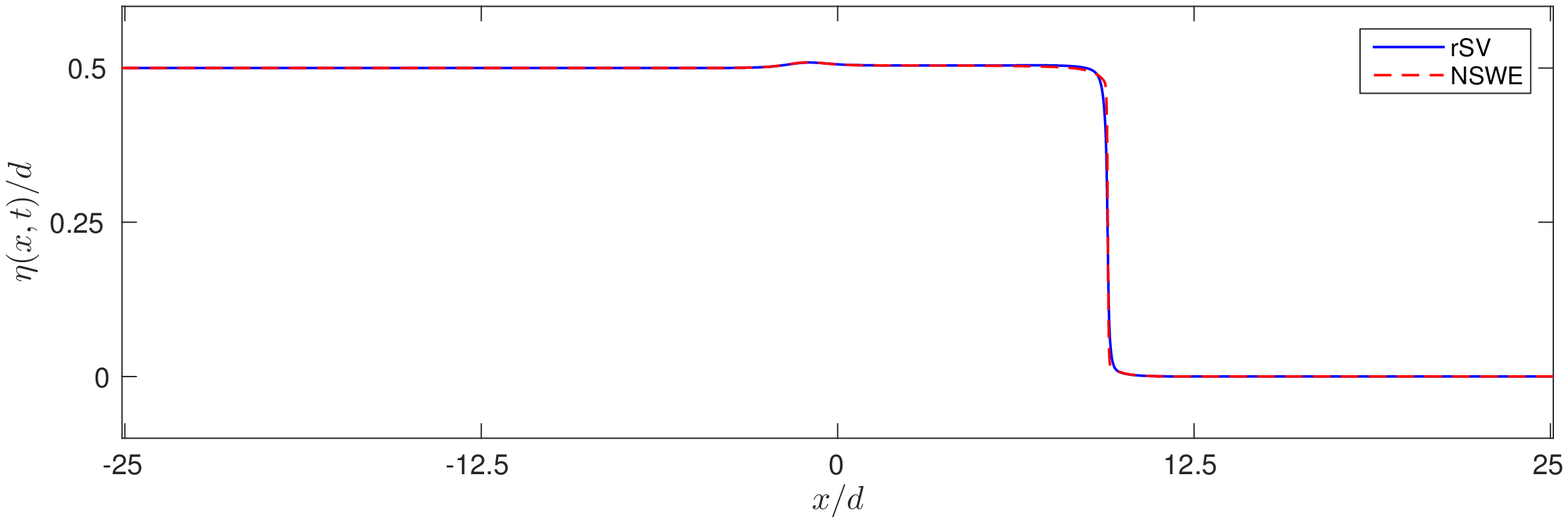}
\caption{\small\em Propagation of a shock wave for the regularisation parameter $\eps\ =\ 10^{-2}\,$. A little wavelet travelling leftwards is due to the fact that the initial condition is chosen to be a smoothed shock wave profile (instead of being a sharp \textsc{Heaviside} function).}
\label{fig:sw}
\end{figure}

The dam-break problem considered above can be slightly refined by choosing thoroughly water levels on both sides of the dam. If it is done according to the classical \textsc{Rankine--Hugoniot} conditions, only one isolated shock wave will emerge to move rightwards with a constant speed, which is also given by the same conditions. Such initial condition parameters are given in Table~\ref{tab:dam} (rightmost column). The shape of the initial condition is the same as above (for the free surface elevation, see Figure~\ref{fig:dam}\textit{a}). The snapshot of the free surface profile at $t\/\sqrt{g/d}\ =\ 5$ is depicted in Figure~\ref{fig:sw}. For the value $\eps\ =\ 10^{-2}$ of the regularisation parameter, we can barely distinguish the two profiles. The most important is that the shock wave position is the same in both models, according to the theoretical predictions (see Section~\ref{sec:rh}). We underline also the fact that there are no oscillations around the jump, showing again that the proposed regularisation is non-dispersive even in fully nonlinear simulations.

\begin{figure}
\centering
\includegraphics[width=0.99\textwidth]{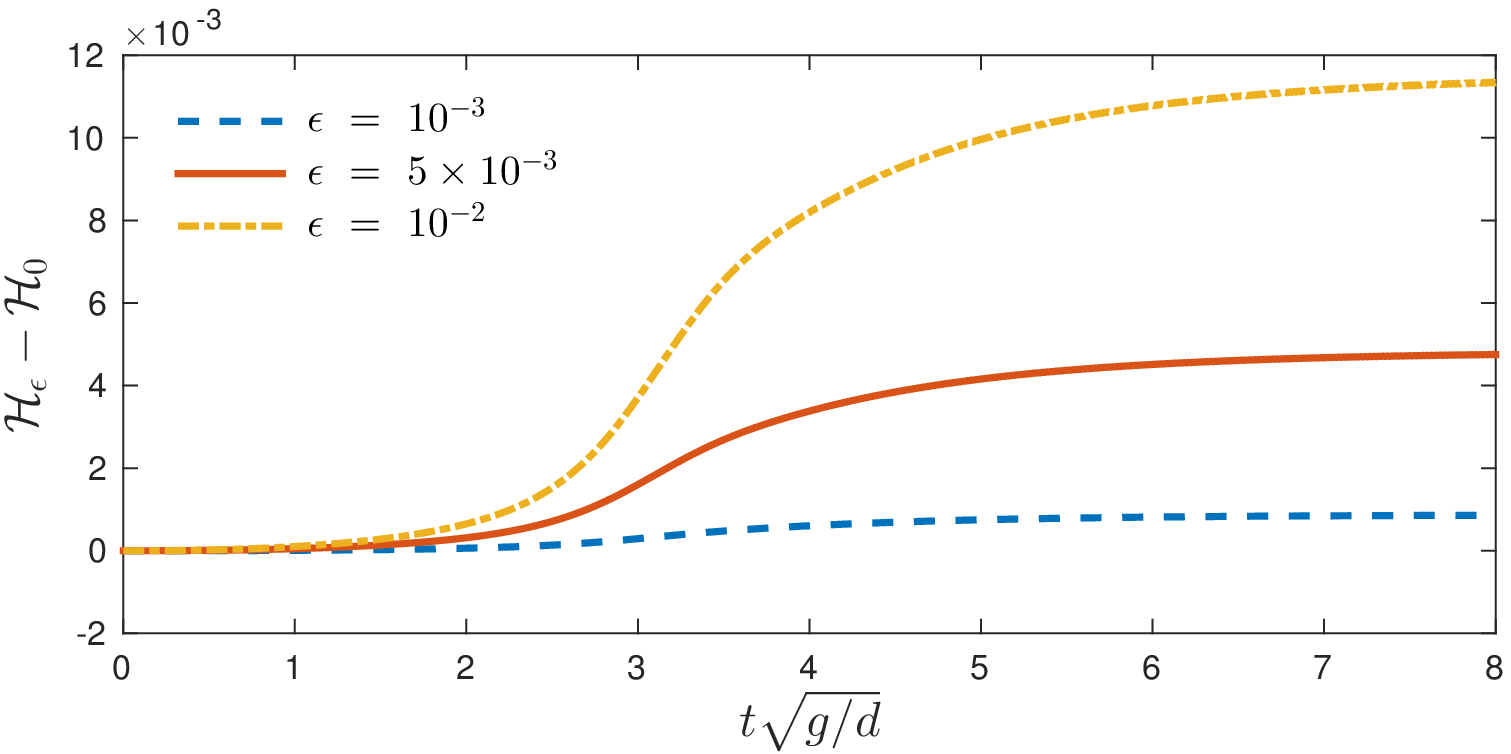}
\caption{\small\em Evolution of the `internal' energy in the formation of a shock.}\label{fig:swene}
\end{figure}

As the wave is stiffening, the `physical' energy $\mathcal{H}_0$ decreases while the `internal' energy $\mathcal{H}_\epsilon\ -\ \mathcal{H}_0$ increases (Fig.~\ref{fig:swene}). The slope being limited in the rSV equations, these variations are plateauing as the steady regularised shock state is reached. This behaviour is expected as it is consistent with the energy jump in the cSV equations. This indicates that the solutions of the rSV equations should tend to the ones of the cSV equations as $\epsilon\ \to\ 0^+\,$. This claim is illustrated numerically but it remains to be proven rigorously.

\section{Discussion}
\label{sec:disc}

This paper presents several new developments for the non-dispersive shallow water equations. First, a new \textsc{Lagrangian} for long wave propagating in shallow water is proposed. This \textsc{Lagrangian} contains two free parameters that can be chosen independently. In the present study, our goal was to remove the frequency dispersion effects in order to obtain a generalised version of the celebrated hyperbolic \textsc{Saint-Venant} equations \cite{SV1871}. Tough the dispersion relation analysis is only linear, the analysis of steady flows shows that the non-dispersive feature remains in nonlinear solutions. In order to investigate the non-dispersive features for unsteady flows, we performed numerical studies of the fully nonlinear equations. The numerical results confirm the absence of dispersive effects in the unsteady nonlinear equations. As a result, we obtained a non-hydrostatic non-dispersive model for nonlinear shallow water waves. Second, for any value of the regularisation parameter $\eps\ >\ 0\,$, we obtain smooth and monotonic solutions to the classical dam-break problem \cite{Dutykh2010c}. The regularising effect is not limited to this peculiar problem. For instance, the hyperbolic shock waves in \textsc{Saint-Venant} equations are replaced by smoothed kink-like fronts \cite{Holden1999} without introducing any dissipative effects into the model. We remind that the rSV model is fully conservative and, by construction, it inherits a variational structure as well. In addition, the main advantage of the method proposed in the present study is that: (i) this regularisation can be supplied with a clear physical meaning; (ii) the whole variational structure of original equations is conserved; (iii) the regularised shock speed is independent of the regularising parameter and is identical to the original speed. To our knowledge, it is the first regularisation which seems to achieve these goals.

The numerical simulations of the regularised equations where performed with a pseudo-spectral scheme. We chose this method for its speed and accuracy but, more importantly, because it requires great regularity of the solution. The fact that it worked fine here demonstrates that other numerical methods should {\em a fortiori} work for the regularised equations. 

An efficient regularisation is obtained only for a well-chosen regularising function $\Rr\,$. \textsc{Leray}-like regularisations are obtained introducing an {\em ad hoc} linear correction. In the present context, this means that we should have used $\Rr\ =\ -d\,u_{\,xt}\ -\ g\,d\,h_{\,xx}$ into the momentum equation, leading to the regularised velocity $\U\/=\/u\/-\/\epsilon\/d^2\/u_{xx}\,$. However, doing so, the speed of the regularised shocks is modified, so this regularisation is not optimal in that respect. Thus, the nonlinear terms in the definition of $\Rr$ play an important role and it is not at all trivial {\em a priori} to find out a regularising term $\Rr$ with all the desired properties. The complexity of the $\Rr$ used here shows that such term would have been very unlikely discovered by trail and error. Thanks to the relaxed variational formulation described in Section~\ref{sec:model}, a suitable choice for $\Rr$ appeared naturally. This is an illustration of the power of this variational approach.   

The proposed rSV equations have to be studied deeper from the mathematical point of view. In particular, the limit of solutions when $\eps\ \to\ 0^+$ has to be rigorously established along the lines of \cite{Lax1983, Hunter1995, Kondo2002a}, to give a few references. Moreover, at the limit, one has to recover not only a weak solution to the original hyperbolic system, but an \emph{entropy} weak solution. More importantly, a question is whether the proposed regularisation allows to obtain existence and uniqueness (or maybe even \emph{stability}) results for the limiting hyperbolic system. We hope that this work will attract the mathematical community's attention to these opportunities.

Moreover, we hope that this approach will be generalised to other important examples of conservation laws which arise in applications \cite{Godlewski1990}. It is clear that the success of this operation will greatly depend on the existence of underlying variational structure of equations.


\subsection*{Acknowledgments}
\addcontentsline{toc}{subsection}{Acknowledgments}

The authors would like to acknowledge the support of CNRS under the PEPS \textsc{InPhyNiTi} 2015 project FARA. D.~\textsc{Dutykh} would like to acknowledge the hospitality of the Laboratory J.~A.~Dieudonn\'e during his visits in 2016.


\bigskip
\addcontentsline{toc}{section}{References}
\bibliographystyle{abbrv}

\bigskip

\end{document}